\documentclass[twocolumn,showpacs,prc,floatfix]{revtex4}
\usepackage{graphicx}
\usepackage[dvips,usenames]{color}
\usepackage{amsmath,amssymb}
\newcommand{\bs}{\begin{sloppypar}} \newcommand{\es}{\end{sloppypar}}
\def\beq{\begin{eqnarray}} \def\eeq{\end{eqnarray}}
\def\beqstar{\begin{eqnarray*}} \def\eeqstar{\end{eqnarray*}}
\newcommand{\bal}{\begin{align}}
\newcommand{\eal}{\end{align}}
\newcommand{\beqe}{\begin{equation}} \newcommand{\eeqe}{\end{equation}}
\newcommand{\p}[1]{(\ref{#1})}

\begin {document}
\title{Spin ordered phase transitions in neutron matter \\ under the presence of
a strong magnetic field
 }
\author{ A. A. Isayev}
%\email{isayev@kipt.kharkov.ua}
 \affiliation{Kharkov Institute of
Physics and Technology, Academicheskaya Street 1,
 Kharkov, 61108, Ukraine
%\\
%Kharkov National University, Svobody Sq., 4, Kharkov, 61077,
%Ukraine
 }
  \author{J. Yang}
 %\email{jyang@ewha.ac.kr}
 \affiliation{Department  of Physics and the Institute for the Early Universe,
 \\
Ewha Womans University, Seoul 120-750, Korea%\\ %\\ Center for High
%Energy Physics, Kyungbook National University, Daegu 702-701,
%Korea}
}
 %\date{\today}
\begin{abstract}
In dense neutron matter under the presence of a strong magnetic
field, considered in the model with the Skyrme effective
interaction, there are possible two types of spin ordered states. In
one of them the majority of neutron spins are aligned opposite to
magnetic field (thermodynamically preferable state), and in other
one the majority of spins are aligned along the field (metastable
state). The equation of state, incompressibility modulus and
velocity of sound are determined in each case with the aim to find
the peculiarities allowing to distinguish between two spin ordered
phases.
\end{abstract}
  \maketitle

\section{Introduction} Magnetars are strongly magnetized neutron stars~\cite{DT}
with emissions powered by the dissipation of magnetic energy.
Magnetars are thought to give the origin to the extremely powerful
short-duration $\gamma$-ray bursts~\cite{HBS,CK}. The magnetic field
strength at the surface of a magnetar is about of
$10^{14}$-$10^{15}$~G~\cite{TD,IShS}. Such huge magnetic fields can
be inferred from observations of magnetar periods and spin-down
rates, or from hydrogen spectral lines.  In the interior of a
magnetar the magnetic field strength may be even larger, reaching
values of about $10^{18}$~G~\cite{CBP,BPL}. Under such
circumstances, the issue of interest is the behavior of
 neutron star matter in a strong magnetic
field~\cite{CBP,BPL,CPL,PG}.

In the recent study~\cite{PG}, neutron star matter was approximated
by  pure neutron matter in a model with the effective nuclear
forces. It was shown that the behavior of spin polarization of
neutron matter in the high density region in a strong magnetic field
crucially depends on whether neutron matter develops a spontaneous
spin polarization (in the absence of a magnetic field) at  several
times  nuclear matter saturation density, or the appearance of a
spontaneous polarization is not allowed  at the relevant densities
(or delayed to much higher densities). The first case  is usual for
the Skyrme forces~\cite{R,S,O,VNB,RPLP,ALP,BPM,KW94,I,IY,RPV,I06},
while the second one is characteristic  for the realistic
nucleon-nucleon (NN) interaction~\cite{PGS,BK,H,VPR,FSS,KS,BB}. In
the former case, a ferromagnetic transition to a totally spin
polarized state occurs while in the latter case a ferromagnetic
transition is excluded at all relevant densities and the spin
polarization remains quite low even in the high density region. If a
spontaneous ferromagnetic transition is allowed,   it was shown in
the subsequent model consideration with the Skyrme effective
forces~\cite{IY09} that the self-consistent equations for the spin
polarization parameter at nonzero magnetic field have not only
solutions corresponding to negative spin polarization (with the
majority of neutron spins oriented opposite to the direction of the
magnetic field) but, because of the strong spin-dependent medium
correlations in the high-density region, also the solutions with
positive spin polarization. In the last case, the formation of a
metastable state with the majority of neutron spins oriented along
the magnetic field is possible in the high-density interior of a
neutron star.

In the present study, we provide the zero-temperature calculations
of the equation of state (EoS), incompressibility modulus and sound
velocity for neutron matter in a strong magnetic field with the aim
to find the peculiarities allowing to distinguish between two
possible spin ordered states - the stable one with negative spin
polarization and the metastable one with positive spin polarization.
It will be shown that in the thermodynamically stable state the
incompressibility modulus and the speed of sound are characterized
by the appearance of the well-defined maximum  just around the
density at which the ferromagnetic (FM) phase transition sets in.
Contrarily to that, such features are missing in the metastable
state. Besides, all calculated quantities behave differently under
changing magnetic field  in stable and metastable states.

At this point, it is worthy to note that  we consider thermodynamic
properties of spin polarized states in neutron  matter in a strong
magnetic field up to the high density region relevant for
astrophysics. Nevertheless, we take into account the nucleon degrees
of freedom only, although other degrees of freedom, such as pions,
hyperons, kaons, or quarks could be important at such high
densities.

% \eject
%\newpage
%\vspace{3mm}
%\begin{center}
\section{Basic equations}
%\textbf{\textsc{2. }}
%\end{center}
Here we only outline the basic equations necessary for further
 calculations, and  a more detailed description  of  a Fermi-liquid
 approach to neutron matter in a strong magnetic field
 can be found in our earlier work~\cite{IY09}.
The normal (nonsuperfluid) states of neutron matter are described
  by the normal distribution function of neutrons $f_{\kappa_1\kappa_2}=\mbox{Tr}\,\varrho
  a^+_{\kappa_2}a_{\kappa_1}$, where
$\kappa\equiv({\bf{p}},\sigma)$, ${\bf p}$ is momentum, $\sigma$ is
the projection of spin on the third axis, and $\varrho$ is the
density matrix of the system~\cite{I,IY,I06}. Further it will be
assumed that the third axis is directed along the external magnetic
field $\bf{H}$.  The self-consistent matrix equation for determining
the distribution function $f$ follows from the minimum condition of
the thermodynamic potential~\cite{AIP} and is
  \begin{eqnarray}
 f=\left\{\mbox{exp}(Y_0\varepsilon+
Y_4)+1\right\}^{-1}\equiv
\left\{\mbox{exp}(Y_0\xi)+1\right\}^{-1}.\label{2}\end{eqnarray}
Here the single particle energy $\varepsilon$ and the quantity $Y_4$
are matrices in the space of $\kappa$ variables, with
$Y_{4\kappa_1\kappa_2}=Y_{4}\delta_{\kappa_1\kappa_2}$, $Y_0=1/T$,
and $ Y_{4}=-\mu_0/T$  being
 the Lagrange multipliers, $\mu_0$ being the chemical
potential of  neutrons, and $T$  the temperature. Given the
possibility for alignment of  neutron spins along or oppositely to
the magnetic field $\bf H$, the normal distribution function of
neutrons and single particle energy can be expanded in the Pauli
matrices $\sigma_i$ in spin
space%~\cite{AIP}
\begin{align} f({\bf p})&= f_{0}({\bf
p})\sigma_0+f_{3}({\bf p})\sigma_3,\label{7.2}\\
\varepsilon({\bf p})&= \varepsilon_{0}({\bf
p})\sigma_0+\varepsilon_{3}({\bf p})\sigma_3.
 \nonumber
\end{align}

Using Eqs.~\p{2} and \p{7.2}, one can express evidently the
distribution functions $f_{0},f_{3}$
 in
terms of the quantities $\varepsilon$: \begin{align}
f_{0}&=\frac{1}{2}\{n(\omega_{+})+n(\omega_{-}) \},\label{2.4}
 \\
f_{3}&=\frac{1}{2}\{n(\omega_{+})-n(\omega_{-})\}.\nonumber
 \end{align} Here $n(\omega)=\{\exp(Y_0\omega)+1\}^{-1}$ and
 \begin{align}
\omega_{\pm}&=\xi_{0}\pm\xi_{3},\label{omega}\\
\xi_{0}&=\varepsilon_{0}-\mu_{0},\;
\xi_{3}=\varepsilon_{3}.\nonumber\end{align}

As follows from the structure of the distribution functions $f$, the
quantities $\omega_{\pm}$ play the role of the quasiparticle
spectrum and  correspond to neutrons with spin up and spin down. The
distribution functions $f$ should satisfy the norma\-lization
conditions
\begin{align} \frac{2}{\cal
V}\sum_{\bf p}f_{0}({\bf p})&=\varrho,\label{3.1}\\
\frac{2}{\cal V}\sum_{\bf p}f_{3}({\bf
p})&=\varrho_\uparrow-\varrho_\downarrow\equiv\Delta\varrho.\label{3.2}
 \end{align}
 Here $\varrho=\varrho_{\uparrow}+\varrho_{\downarrow}$ is the total density of
 neutron matter, $\varrho_{\uparrow}$ and $\varrho_{\downarrow}$  are the neutron number densities
 with spin up and spin down,
 respectively. The
quantity $\Delta\varrho$  may be regarded as the neutron spin order
parameter. The spin ordering in neutron matter can also be
characterized by the  neutron spin
polarization parameter $$ %\beqe
\Pi=\frac{\varrho_{\uparrow}-\varrho_{\downarrow}}{\varrho}\equiv\frac{\Delta\varrho}{\varrho}.
$$
The spin order parameter  determines the magnetization of the system
$M=\mu_n \Delta\varrho$, $\mu_n$ being the neutron magnetic moment.
The magnetization may contribute to the internal magnetic field
$B=H+4\pi M$. However, we will assume, analogously to
Refs.~\cite{BPL,PG}, that the contribution of the magnetization
 to the magnetic field
$B$ remains small for all relevant densities and magnetic field
strengths, and, hence, $B\approx H$. This assumption holds true due
to the tiny value of the neutron magnetic moment
$\mu_n=-1.9130427(5)\mu_N\approx-6.031\cdot10^{-18}$ MeV/G~\cite{A}
($\mu_N$ being the nuclear magneton)
 and is confirmed numerically in a subsequent integration of the self-consistent equations.

In order to get the self--consistent equations for the components of
the single particle energy, one has to set the energy functional of
the system. In view of the above approximation, it reads~\cite{IY}
\begin{align} E(f)&=E_0(f,H)+E_{int}(f)+E_{field},\label{enfunc} \\
{E}_0(f,H)&=2\sum\limits_{ \bf p}^{}
\underline{\varepsilon}_{\,0}({\bf p})f_{0}({\bf p})-2\mu_n
H\sum\limits_{ \bf p}^{} f_{3}({\bf p}),\nonumber
\\ {E}_{int}(f)&=\sum\limits_{ \bf p}^{}\{
\tilde\varepsilon_{0}({\bf p})f_{0}({\bf p})+
\tilde\varepsilon_{3}({\bf p})f_{3}({\bf p})\},\nonumber\\
E_{field}&=\frac{H^2}{8\pi}\cal V,\nonumber\end{align} where
\begin{align}\tilde\varepsilon_{0}({\bf p})&=\frac{1}{2\cal
V}\sum_{\bf q}U_0^n({\bf k})f_{0}({\bf
q}),\;{\bf k}=\frac{{\bf p}-{\bf q}}{2}, \label{flenergies}\\
\tilde\varepsilon_{3}({\bf p})&=\frac{1}{2\cal V}\sum_{\bf
q}U_1^n({\bf k})f_{3}({\bf q}). %\nonumber
\end{align}
Here  $\underline\varepsilon_{\,0}({\bf p})=\frac{{\bf
p}^{\,2}}{2m_{0}}$ is the free single particle spectrum, $m_0$ is
the bare mass of a neutron, $U_0^n({\bf k}), U_1^n({\bf k})$ are the
normal Fermi liquid (FL) amplitudes, and
$\tilde\varepsilon_{0},\tilde\varepsilon_{3}$ are the FL corrections
to the free single particle spectrum. Note that the field
contribution $E_{field}$, being the energy of the magnetic field in
the absence of matter, leads only to the constant shift of the total
energy and, by this reason, can be omitted.  Using Eq.~\p{enfunc},
one can get the self-consistent equations in the form~\cite{IY}
\begin{align}\xi_{0}({\bf p})&=\underline\varepsilon_{\,0}({\bf
p})+\tilde\varepsilon_{0}({\bf p})-\mu_0, \label{14.2}\\
\xi_{3}({\bf p})&=-\mu_nH+\tilde\varepsilon_{3}({\bf
p}).\label{14.3}
\end{align}

 To obtain
 numerical results, we  utilize the  effective Skyrme interaction~\cite{VB}.
 Expressions for the normal FL amplitudes in terms of the Skyrme
  force parameters were written in Refs.~\cite{AIP,IY3}.
Thus, using expressions~\p{2.4} for the distribution functions $f$,
we obtain the self-consistent equations~\p{14.2}, \p{14.3} for the
components of the single-particle energy $\xi_{0}({\bf p})$ and
$\xi_{3}({\bf p})$, which should be solved jointly with the
normalization conditions~\p{3.1}, \p{3.2}. Further we do not take
into account the effective tensor forces, which lead to coupling of
the momentum and spin degrees of freedom, and, correspondingly, to
anisotropy in the momentum dependence of FL amplitudes with respect
to the spin quantization axis.

If the self-consistent equations have a few branches of the
solutions, it is necessary to compare the corresponding energies (at
zero temperature) in order to decide which solution is
thermodynamically preferable. The energy per neutron, $E/A$, can be
directly calculated from Eq.~\p{enfunc}. The equation of state (EoS)
of neutron matter in a strong magnetic field then can be obtained
from the equation
\begin{equation}\label{press}
P=\varrho^2\frac{\partial\,\bigl ( e/\varrho \bigr
)}{\partial\,\varrho}
\end{equation}
where $e=\varrho(mc^2+E/A)$ is the energy density, which includes
also the rest energy term. The incompressibility modulus,
$K=9\frac{\partial P}{\partial \varrho}$, according to
Eq.~\p{press}, reads
\begin{equation}\label{compress}
K=9\varrho^2\frac{\partial\,^2\,\bigl ( E/A \bigr
)}{\partial\varrho^2}+18\frac{P}{\varrho}.
\end{equation}
The speed of sound, $v_s=c\sqrt{\frac{\partial P}{\partial e}}$, can
be related to the incompressibility modulus by the equation
\begin{equation}\label{vs}
    v_s=c\sqrt{\frac{K}{9(mc^2+\frac{E}{A}+\frac{P}{\varrho})}}\,.
\end{equation}

%\begin{center}
%\textbf{\textsc{3. EOS  OF DENSE NEUTRON MATTER IN A STRONG MAGNETIC
%FIELD}}
%\end{center}
\section{EoS  of dense neutron matter in a strong magnetic
field}

The self-consistent equations were analyzed at zero temperature in
Ref.~\cite{IY09} for the magnetic field strengths up to
$H_{max}\sim10^{18}$~G, allowed by a scalar virial
theorem~\cite{LS}, in the model consideration with SLy4 and SLy7
Skyrme effective forces~\cite{CBH}.  It was shown that a
thermodynamically stable branch of solutions for the spin
polarization parameter as a function of density corresponds to
negative spin polarization when the majority of neutron spins are
oriented opposite to the direction of the magnetic field. Besides,
beginning from some threshold density $\varrho_{th}\sim4\varrho_0$,
being slightly dependent on the magnetic field strength,  the state
with positive spin polarization can also be realized as a metastable
state in neutron matter (cf. branches $\Pi_1$ and $\Pi_3$ in Fig.~2
of Ref.~\cite{IY09}). This conclusion was based on the comparison of
the free energies of two states which turn out to be very close to
each other~\cite{IY09,IY10}. However, as it will  be shown later,
additional constraints, such as, e.g., stability of the system with
respect to the density fluctuations, will define more accurately the
density range admissible for the state with positive spin
polarization.

In this work, the previous study~\cite{IY09} will be extended by
calculating the EoS of dense neutron matter in a strong magnetic
field for various branches of solutions of the self-consistent
equations. Each possible state should match the constraint $K>0$ for
allowable densities and magnetic field strengths being the condition
of the mechanical stability of the system. Besides, in the
high-density region, the velocity of sound should not exceed the
speed of light in the vacuum, $v_s<c$. Note that further the
contribution of the magnetic field pressure to the total pressure
will be omitted because in the magnetic fields up to $10^{18}$~G the
magnetic field pressure is still small compared to the matter
pressure in the high-density region of interest.

\begin{figure}[tb]
%\begin{figure}[p]
\begin{center}
\includegraphics[width=8.6cm,keepaspectratio]{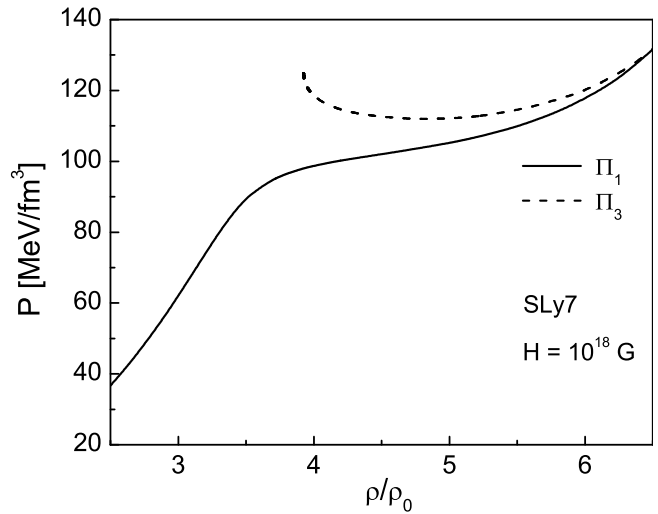}
\end{center}
\vspace{-2ex} \caption{Pressure vs. density for the branches $\Pi_1$
(stable) and $\Pi_3$ (metastable) of spin polarization in neutron
matter with the Skyrme SLy7 interaction at $H=10^{18}$~G.}
\label{fig1}\vspace{-0ex}
\end{figure}

First, we present the results of determining the zero-temperature
EoS of neutron matter in  a strong magnetic field at the density
region where both stable and metastable spin ordered states
 can be realized. Because the results
of calculations with SLy4 and SLy7 Skyrme forces are very close,
here we present the obtained dependences only for the SLy7 Skyrme
interaction. Fig.~1 shows the pressure of neutron matter as a
function of density for two branches of spin polarization, stable
$\Pi_1$ and metastable $\Pi_3$, corresponding to negative and
positive  polarizations, respectively (the branch $\Pi_2$ with
positive spin polarization considered in Ref.~\cite{IY09} has the
considerably larger energy per neutron as compared to the previous
ones). For the branch $\Pi_1$, the pressure is the increasing
function of the density for all relevant densities, and, hence, the
incompressibility coefficient is always positive. However, for the
branch $\Pi_3$, beginning from the threshold density
$\varrho_{th}\approx3.92\varrho_0$ up to the density
$\varrho_c\approx4.85\varrho_0$ (at $H=10^{18}$~G), the pressure
decreases with the density. Hence, in this density range the
incompressibility coefficient is negative and the metastable state
characterized by the branch $\Pi_3$ of positive spin polarization
cannot appear at these densities. However, beyond the critical
density $\varrho_c$, the metastable state with  positive spin
polarization is allowed by the criterion  $K>0$. Note that the EoS
for the metastable state of neutron matter in a strong magnetic
field is stiffer than that for the thermodynamically equilibrium
state.

\begin{figure}[tb]
%\begin{figure}[p]
\begin{center}
\includegraphics[width=8.6cm,keepaspectratio]{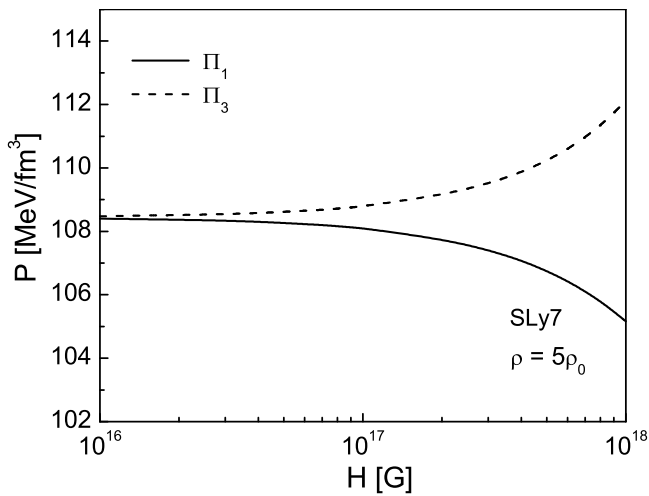}
\end{center}
\vspace{-2ex} \caption{Pressure vs. magnetic field strength for the
branches $\Pi_1$  and $\Pi_3$  of spin polarization in neutron
matter with the Skyrme SLy7 interaction at $\varrho=5\varrho_0$.}
\label{fig2}\vspace{-0ex}
\end{figure}

Fig.~2 shows the pressure of neutron matter as a function of the
magnetic field strength  for the branches $\Pi_1$ and $\Pi_3$
 of spin polarization at $\varrho=5\varrho_0$.  It is seen that the dependence of
 the EoS for stable and metastable
branches is different: for the branch $\Pi_1$ the EoS becomes softer
with the magnetic field while for the branch $\Pi_3$ stiffer. These
calculations  show that the impact of the magnetic field on the EoS
remains small up to  the field strengths of about $10^{17}$~G.

Fig.~3 shows the zero-temperature incompressibility modulus of
neutron matter in a strong magnetic field  as a function of density
for the branches $\Pi_1$ and $\Pi_3$ of spin polarization. For the
branch $\Pi_3$,  the incompressibility modulus monotonously
increases with the density and changes sign from negative to
positive at the critical  density $\varrho_c$, marking the stability
range with respect to density fluctuations at densities beyond
$\varrho_c$. As a consequence, if the metastable state with positive
spin polarization can be realized  in the high-density region of
neutron matter in a strong magnetic field, under decreasing density
(going from the interior to the outer regions of a magnetar) it
changes at the critical density $\varrho_c$ to a thermodynamically
stable state with negative spin polarization.

For the branch $\Pi_1$, the behavior of incompressibility modulus is
nonmonotone. The most important peculiarity is that just around the
density ($\varrho_{tr}\approx 3.16\varrho_0$ at $H=10^{18}$~G) at
which the magnitude of the spin polarization parameter for the
branch $\Pi_1$ begins rapidly to increase (cf. Fig.~2 of
Ref.~\cite{IY09}), the increasing behavior of the incompressibility
modulus with the density changes on the decreasing one. Because the
density $\varrho_{tr}$ can be regarded as the density at which  a
ferromagnetic state  sets in, this qualitative feature in the
behavior of the incompressibility modulus can be used as the
characteristic of the density-driven FM phase transition in neutron
matter possessing equilibrium spin polarization. The
incompressibility modulus decreases till the density about
$4\varrho_0$ at which the spin polarization parameter is well
developed and gets about two third of its strength. Then there is
the plateau in the density dependence of incompressibility modulus
till the density about $5\varrho_0$ beyond which the
incompressibility modulus begins gradually to increase. Note that
the noticeable  decrease of the incompressibility modulus around the
 density of the transition to the ferromagnetic state was mentioned also
in Ref.~\cite{PGNP}, although the total incompressibility modulus
was not explicitly shown there, but only that for the spin-up and
spin-down neutron components in the state with equilibrium spin
polarization. Besides, in Refs.~\cite{PG,PGNP}, there were no any
calculations  related to the metastable branch of spin polarization
because this branch itself was missed in these studies.

\begin{figure}[tb]
%\begin{figure}[p]
\begin{center}
\includegraphics[width=8.6cm,keepaspectratio]{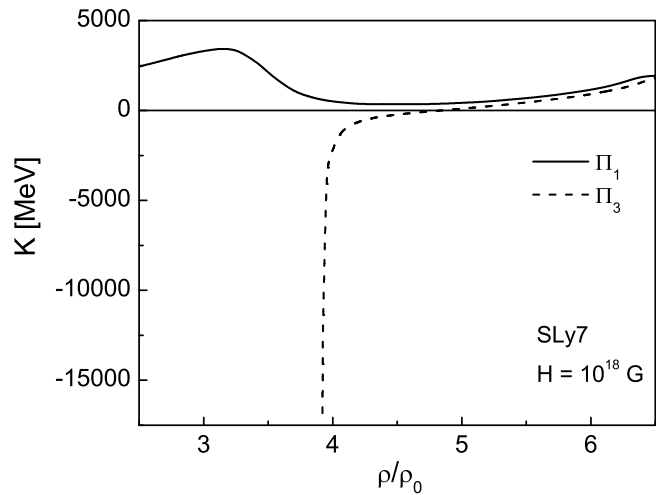}
\end{center}
\vspace{-2ex} \caption{Same as in Fig.~1, but for the
incompressibility modulus of neutron matter in a strong magnetic
field. } \label{fig3}\vspace{-0ex}
\end{figure}

Fig.~4 shows the incompressibility modulus of neutron matter   as a
function of the magnetic field strength for the branches $\Pi_1$ and
$\Pi_3$ of spin polarization at the density $\varrho=5\varrho_0$.
For the branch $\Pi_1$, the incompressibility modulus increases with
the magnetic field strength while for the branch $\Pi_3$ it
decreases. It follows from these calculations  that the impact of
the magnetic field on the incompressibility modulus remains mild up
to the field strengths of about $10^{17}$~G.

\begin{figure}[tb]
%\begin{figure}[p]
\begin{center}
\includegraphics[width=8.6cm,keepaspectratio]{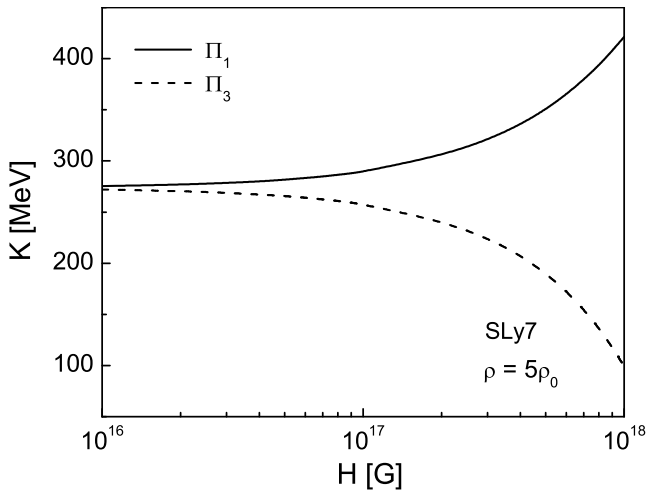}
\end{center}
\vspace{-2ex} \caption{Incompressibility modulus vs. magnetic field
strength for the branches $\Pi_1$  and $\Pi_3$  of spin polarization
in neutron matter with the Skyrme SLy7 interaction at
$\varrho=5\varrho_0$.} \label{fig4}\vspace{-0ex}
\end{figure}

  Fig.~5 shows the sound velocity in neutron matter under
the presence of a strong magnetic field  as a function of density
for the branches $\Pi_1$ and $\Pi_3$ of spin polarization. For the
branch $\Pi_3$, Eq.~\p{vs} automatically guarantees the fulfillment
of the condition $K>0$ (for all relevant densities $E/A>0$). It is
seen that for both branches at the relevant densities the
superluminous regime doesn't occur. While for the branch $\Pi_3$ the
sound velocity monotonously increases with the density, for the
branch $\Pi_1$ it has non-monotone behavior. In fact, near the
transition density $\varrho_{tr}$ the sound velocity in a
thermodynamically stable state has a clear peak structure
considerably decreasing at the densities where the ferromagnetic
phase sets in. This feature, together with the presence of the
maximum in the density dependence of the incompressibility modulus,
can be used for the identification of the density-driven FM phase
transition in neutron matter possessing equilibrium spin
polarization. On the other hand, the incompressibility modulus and
the speed of sound monotonously increase with density in the
metastable state with positive spin polarization, and hence, these
features can be used for distinguishing between the
thermodynamically stable (negative spin polarization) and metastable
(positive spin polarization) states in neutron matter under the
presence of a strong magnetic field.

\begin{figure}[tb]
%\begin{figure}[p]
\begin{center}
\includegraphics[width=8.6cm,keepaspectratio]{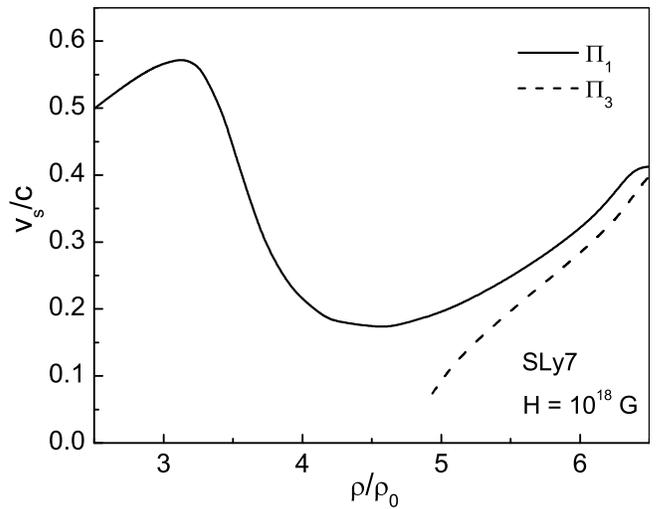}
\end{center}
\vspace{-0ex} \caption{Same as in Fig.~1, but for the sound velocity
in neutron matter under the presence of a strong magnetic
field.} \label{fig5}%\vspace{0ex}
\end{figure}

Fig.~6 shows the sound velocity in neutron matter   as a function of
the magnetic field strength for the branches $\Pi_1$ and $\Pi_3$ of
spin polarization at the density $\varrho=5\varrho_0$.  For the
branch $\Pi_1$, the sound velocity  increases with the magnetic
field strength while for the branch $\Pi_3$ it decreases. These
trends are quite similar to those in the behavior of the
incompressibility modulus $K(H)$ for the branches $\Pi_1$ and
$\Pi_3$.

\begin{figure}[tb]
%\begin{figure}[p]
\begin{center}
\includegraphics[width=8.6cm,keepaspectratio]{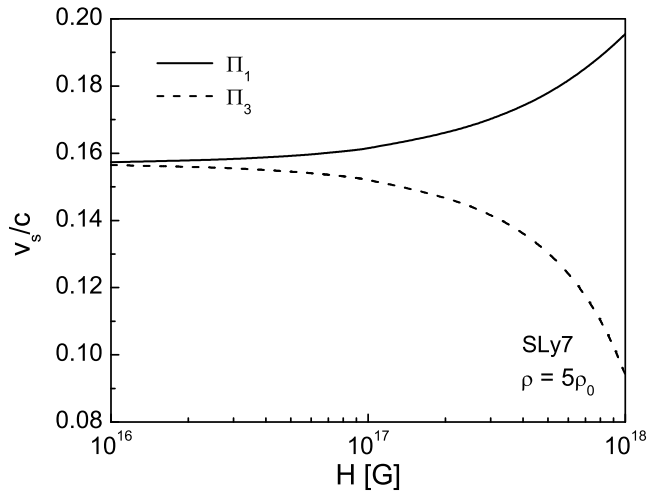}
\end{center}
\vspace{-0ex} \caption{Same as in Fig.~4 but for the sound velocity
in neutron matter under the presence of a strong magnetic
field. } \label{fig6}%\vspace{0ex}
\end{figure}

\section{Conclusions}

In dense neutron matter under the presence of a strong magnetic
field, considered in the model with the Skyrme effective
interaction, there are possible two types of spin ordered states:
the one with the majority of neutron spins aligned opposite to
magnetic field (thermodynamically preferable state), and the other
one with the majority of spins aligned along the field (metastable
state). The equation of state, incompressibility modulus and
velocity of sound have been determined in each case for SLy7 Skyrme
force with the aim to find the peculiarities allowing to distinguish
between two spin ordered phases.

For the stable state with the branch $\Pi_1$ of negative spin
polarization, the EoS is softer than that for metastable state with
the branch $\Pi_3$ of positive spin polarization. The condition of
the positiveness of the incompressibility modulus, $K>0$, is
satisfied for all relevant densities and magnetic field strengths
for the stable branch $\Pi_1$. However, for the branch $\Pi_3$,
although formally the solutions of the self-consistent equations
exist at densities larger than some threshold one, $\varrho_{th}$,
the condition $K>0$ is satisfied only at the densities larger than
the critical one, $\varrho_c$ (e.g., for $H=10^{18}$~G,
$\varrho_{th}\approx3.92\varrho_0$ and
$\varrho_c\approx4.85\varrho_0$).   As a consequence, if the
metastable state with positive spin polarization can be realized  in
the high-density region of neutron matter in a strong magnetic
field, under decreasing density (going from the interior to the
outer regions of a magnetar) it changes at the critical density
$\varrho_c$ to a thermodynamically stable state with negative spin
polarization.

For the thermodynamically stable branch $\Pi_1$, the
incompressibility modulus and the speed of sound are characterized
by the appearance of the well-defined maximum just around the
density at which the ferromagnetic phase sets in. The last
qualitative features can be used for the identification of the
density-driven FM phase transition in neutron matter, possessing
equilibrium spin polarization, under the presence of a strong
magnetic field. Contrarily to the previous case, for the branch
$\Pi_3$ of positive spin polarization, the incompressibility modulus
and the speed of sound monotonously increase with density that can
be used to distinguish between two different spin ordered phases.

The dependence of all calculated quantities on the magnetic field
strength $H$ turns out to be different for two spin ordered phases.
For the thermodynamically stable branch $\Pi_1$, the
incompressibility modulus and sound velocity increase with $H$ while
the pressure
 decreases. The exactly opposite tendency  has been found for the branch
 $\Pi_3$ of positive spin polarization that also allows one to differentiate between
 spin
 polarized states with opposite polarizations in neutron matter under
 the presence of a strong magnetic field.

As yet one problem for consideration, it would be  interesting to
study the role of finite temperature effects  on the EoS,
incompressibility modulus and speed of sound in  dense neutron
matter in a strong magnetic field.  As has been shown already, these
effects can lead to a number of nontrivial features such as, e.g.,
unusual behavior of the entropy in various spin ordered
systems~\cite{IY10,IY2,I3,I4}.

J.Y. was supported by grant 2010-0011378 from Basic Science Research
Program through NRF of Korea funded by MEST and by  grant R32-10130
from WCU project of MEST and NRF.

 %\vspace{3mm}
%--------------------   Bibliography  ------------------------------------%
%\begin{center}

%\end{center}
%\end{multicols}
\end{document}